\documentclass[reprint, amsmath,amssymb, aps]{revtex4-1}

\usepackage{graphicx}
\usepackage{dcolumn}
\usepackage{bm}
\usepackage{amsmath}

\begin{document}

\title{Huge permittivity and premature metallicity in Bi$_2$O$_2$Se single crystals}


\author{Zhuokai Xu $^{1,2,3}$, Jialu Wang$^{1,2,3}$, Tao Wang$^{2,3}$,  Wanghua Hu$^{2,3}$, Xiaohui Yang$^{2,3}$, and  Xiao Lin$^{2,3}$ }\email{E-mail address: linxiao@westlake.edu.cn}

\affiliation{$^{1}$  Department of Physics, Zhejiang University, Hangzhou 310027, Zhejiang Province, China\\
$^{2}$  Key Laboratory for Quantum Materials of Zhejiang Province, School of Science, Westlake University, 18 Shilongshan Road, Hangzhou, Zhejiang 310024, China \\
$^{3}$	Institute of Natural Sciences, Westlake Institute for Advanced Study, 18 Shilongshan Road, Hangzhou 310024, Zhejiang Province, China \\
}

\date{\today}



\begin{abstract}
Bi$_2$O$_2$Se is a promising material for next-generation semiconducting electronics. It exhibits premature metallicity on the introduction of a tiny amount of electrons, the physics behind which remains elusive. Here we report on transport and dielectric measurements in Bi$_2$O$_2$Se single crystals at various carrier densities. The temperature-dependent resistivity ($\rho$) indicates a smooth evolution from the semiconducting to the metallic state. The critical concentration for the metal-insulator transition (MIT) to occur is extraordinarily low ($n_\textrm{c}\sim10^{16}$ cm$^{-3}$). The relative permittivity of the insulating sample is huge ($\epsilon_\textrm{r}\approx155(10)$) and varies slowly with temperature. Combined with the light effective mass, a long effective Bohr radius ($a_\textrm{B}^*\approx36(2)$ $\textrm{nm}$) is derived, which provides a reasonable interpretation of the metallic prematurity according to Mott's criterion for MITs. The high  electron mobility ($\mu$) at low temperatures may result from the screening of ionized scattering centers due to the huge $\epsilon_\textrm{r}$. Our findings shed light on the electron dynamics in two dimensional (2D) Bi$_2$O$_2$Se devices.



\end{abstract}

\maketitle

\section{Introduction}\label{sec:1}
~\\
Bi$_2$O$_2$Se (BOS) is a layered semiconductor with a tetragonal crystal structure~\cite{Boller1973} and is a promising candidate as a next-generation low-power, high-performance semiconducting material. BOS has a robust band gap ($\Delta=0.8$ eV), high air-stability, and most importantly, an unexpectedly high electron mobility ($\mu\approx450$ cm$^2$V$^{-1}$s$^{-1}$) at room temperature (room-$T$)~\cite{Wu2017,Chen2018}, surpassing most functional 2D semiconductors, including MoS$_2$~\cite{Radisavljevic2011}. Moreover, BOS was reported to exhibit record-breaking optoelectronic performance~\cite{Fu2018}.

A search of the literature revealed intensive studies on the functionality of Bi$_2$O$_2$Se thin films~\cite{Sun2020}, in the field of electronics~\cite{Zhang2018,Tan2019,LiTR2020,Zhang2019,Qu2020}, flexible electronics~\cite{ZhangC2020}, optoelectronics~\cite{Wu2017-2, Fu2018, LiJ2018, Khan2019, Tong2019}, and thermoelectrics~\cite{Pan2019}, while fundamental transport researches remain in paucity~\cite{Meng2018,Meng2019,Wang2020,Li2020}. Meng \emph{et al.} detected a crossover between weak antilocalization and weak localization by electrostatically tuning the carrier density~\cite{Meng2018,Meng2019}, indicating a strong spin-orbital coupling in BOS nanoplates. A more recent study by parts of the authors reported the $T$-square resistivity in BOS cannot be explained by interband electron-electron scattering and Umklapp events, implying the absence of a proper understanding of the ubiquitous $T^2$-resistivity in Fermi liquids~\cite{Wang2020}.

A few other fundamental transport properties also remain elusive. For instance, an urgent work is to investigate MIT in BOS, which was one of the central issues with silicon and germanium decades ago~\cite{Thomas1982,Rosenbaum1983}. It is relevant to the following mysteries. How can the metallic phase survive at such low electron concentrations as $n\sim10^{17}$ cm$^{-3}$~\cite{Wang2020}? Why is the low-temperature (low-$T$) $\mu$ unexpectedly high $\sim300000$ cm$^2$V$^{-1}$s$^{-1}$~\cite{Chen2018,Lv2019,Wang2020} and comparable to that of the best topological semimetals~\cite{Liang2015,Huang2015}, given the conventional parabolic dispersion of the conducting band in BOS~\cite{Wu2017,Wang2020}?

These questions could not be addressed previously as the reported BOS single crystals always appeared to be metallic due to the nonstoichiometry of the as-grown materials~\cite{Wu2017,Wang2020,Tong2018,Lv2019,Wu2019}. In this context, density functional theory (DFT) calculations by Fu \emph{et al.} suggested that electrons are spontaneously ionized from donor sites and that the impurity levels lie above the lowest conduction band minimum~\cite{YanBH2018}.

Here, we synthesize bulk insulating BOS single crystals for the first time, enabling us to deal with the aforementioned questions. We find MIT occurs at extraordinarily low electron densities ($n_\textrm{c}\sim 10^{16}$ cm$^{-3}$). The permittivity ($\epsilon$) in insulating samples is huge, amounting to $155(10)\epsilon_0$. Applying the hydrogen model to the semiconductors, we derive a long effective Bohr radius ($a_\textrm{B}^*\approx36(2)$ $\textrm{nm}$). We conclude that the premature metallic phase in BOS is a direct consequence of the long $a_\textrm{B}^*$, in accordance with Mott's criterion for metal-insulator transitions~\cite{Mott1990}. The high $\mu$ at low-$T$ is interpreted in terms of the large effective screening of the ionized impurity scattering centers. Our findings will be of great interest to the community studying the functionality of BOS devices.

\begin{figure}
{\includegraphics[width=8cm]{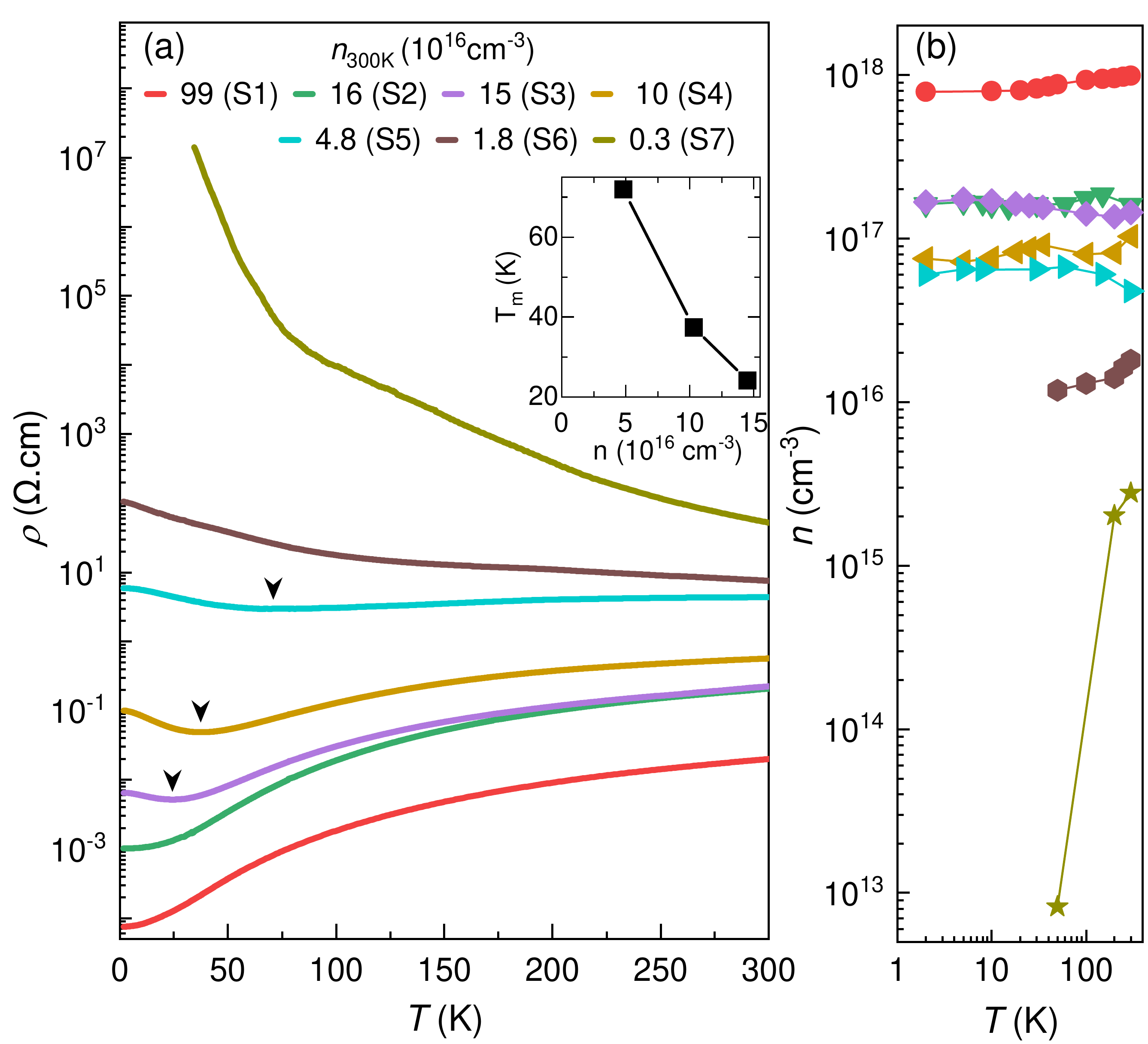}}
\caption{Resistivity ($\rho$) and Hall carrier concentration ($n$) for various BOS single crystals. a) Temperature dependence of the
  resistivity in a semi-log plot. The arrows mark the temperature ($T_\textrm{m}$) at which $\rho$ has a minimum. The inset plots $T_\textrm{m}$ as a function of $n$. b) $n$ as a function of temperature. For S6, $n$ is only presented above 50 K, below which the Hall signal is screened by the large longitudinal signal during the measurement because of the mismatch of the Hall contacts.}
\end{figure}

\section{Materials and method}\label{sec:2}

BOS single crystals were grown by a chemical vapor transport method (CVT), using poly-crystalline powders as precursors. To obtain polycrystals, thoroughly mixed Bi (5N), Se (5N), and Bi$_2$O$_3$ (5N) powders were pressed into pellets and sealed in a quartz tube. The tube was heated at 573 K for 6 h and then at 773 K for a further 12 h. The obtained products were reground. Around 10 $\textrm{g}$ of the reground powder were sealed in a new quartz tube and placed in a horizontal furnace with a temperature gradient between 1123 K and 1023 K over one week. The quality of the as-grown single crystals is as good as previously reported in ref~\cite{Wang2020}. Detailed characterization of the samples can be found in the supplementary information.

X-ray diffraction patterns were established using a Bruker D8 Advance X-ray diffractometer, using the Cu K$ \alpha$ radiation line, at room temperature. The composition of samples was determined by an energy-dispersive X-ray spectrometer (EDX) attached to a Zeiss field emission scanning electron microscope (SEM). The transport measurements were performed using a standard four-terminal method, with either a Keithley 2000 or Keithley 6517B as the voltmeter and a Keithley 6211 as the current source. The dielectric properties were studied using a Hioki IM3536 LCR meter. The temperature-dependent measurements were done in an Oxford Teslatron-PT equipped with a 14 T magnet and with a temperature range from 1.6 K to 300 K. Ohmic contacts were obtained by sputtering gold onto the samples. The basic transport parameters for seven samples are summarized in Table I.

\begin{table*}
  \caption{Basic transport parameters for seven BOS samples. $n_\textrm{300K}$ is the Hall concentration measured at 300 K. $\mu_\textrm{2K}$ is the Hall mobility at 2 K. RRR is the residual-resistivity ratio. $\rho_\textrm{2K}$ is the resistivity at 2 K.}
\begin{tabular}{|c|c|c|c|c|c|c|c|c|}
  \hline
  & 1 & 2 & 3 & 4 & 5 & 6 & 7 \\

 \hline
  $n_\textrm{300K}$ (cm$^{-3}$) & $9.9\times10^{17}$ & $1.6\times10^{17}$ & $1.5\times10^{17}$ &$1\times10^{17}$ & $4.8\times10^{16}$ & $1.8\times10^{16}$ & $3\times10^{15}$ \\

 \hline

  $\mu_\textrm{2K}$ (cm$2$V$^{-1}$s$^{-1}$) & 86363 & 38127 & 5761 & 827 & 26 & 5.08 $^{1)}$ & -- \\

  \hline
  RRR &267 & 205 & 35 & 5.7 & 0.74 & 0.072 & -- \\

 \hline

  $\rho_\textrm{2K}$ ($\Omega.\textrm{cm}$) & $7.49\times10^{-5}$ & $1.02\times10^{-3}$ & $6.48\times10^{-3}$ & 0.1 & 5.92 & 105.2 & $5\times10^{7}$~$^{2)}$ \\

  \hline
\end{tabular}
\begin{tabular}{ccc}
\multicolumn{3}{l}{$^{1)}$ $\mu$ is calculated from the Hall concentration measured at 50 K.}\\
\multicolumn{3}{l}{$^{2)}$ The value is taken at 35 K, below which $\rho$ diverges.} \\
\multicolumn{3}{l}{-- ~~refers to quantities which could not be determined.}\\
\end{tabular}

\end{table*}

Fig. 1a presents the temperature dependence of the resistivity ($\rho$) from 2 K to 300 K for seven BOS samples with $n$ differing by two orders of magnitude. The temperature-dependent Hall concentration ($n$) is plotted in Fig. 1b, in which $n$ varies slowly with temperature ($T$) for most of the samples except for sample 7 (S7) with the lowest $n$. From the $\rho-T$ profile, we define three regions.

In region-I (R-I), containing S1 and S2, with $n$ above a threshold concentration ($n^*\sim10^{17}$ cm$^{-3}$), $\rho$ declines monotonically with reducing temperature. $\mu_\textrm{2K}$ and the residual-resistivity ratio (RRR) above $n^*$ are large ($\mu_\textrm{2K}\sim10^5$ cm$^2$V$^{-1}$s$^{-1}$ and $\textrm{RRR}\sim300$) as seen in Fig. 2a and b. There is no doubt that the ground state in R-I is metallic. One may refer to ref~\cite{Wang2020} for quantum oscillation measurements on BOS with similar $n$, which implies the formation of Fermi surface, a more stringent definition for a metallic state~\cite{Mackintosh1963}.

\begin{figure}
  \includegraphics[width=8cm]{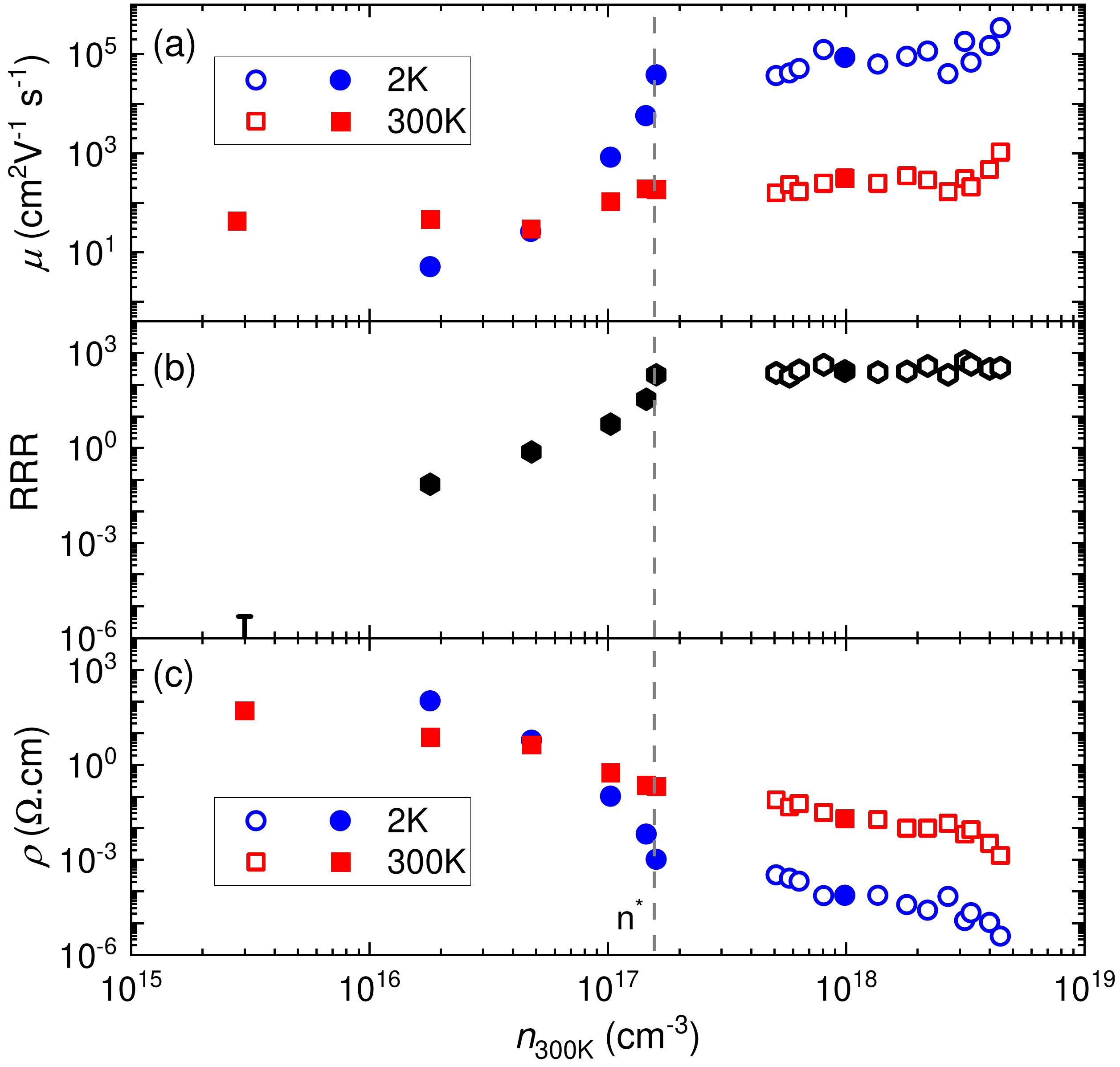}
  \caption{Transport parameters as a function of the Hall carrier concentration at 300 K for various samples. a) Electron mobility ($\mu$) at 2K and 300K. b) Residual-resistivity ratio (RRR) calculated from $\rho_\textrm{300K}/\rho_\textrm{2K}$. RRR for S7 is from the ratio between $\rho_\textrm{300K}$ and the highest detectable resistivity measured at around 35 K. Hence, it only represents the upper bound. c) Resistivity at 2 K and 300 K. The open points are from ref~\cite{Wang2020}. The dashed line marks the threshold concentration ($n^*$) below which the localization effects appear.}
  \label{Fig2}
\end{figure}

We then place S3-S6 in R-II, where $n$ is below $n^*$. For S3-S5, the evolution of $\rho$ is nonmonotonic. $\rho$ is metal-like above a characteristic temperature ($T_\textrm{m}$), but shows an upturn below. $T_\textrm{m}$ increases with reducing $n$, as seen in the inset of Fig. 1a. For S6, though $\rho$ increases with declining $T$ across the full $T$-range, we will see below that the zero-$T$ resistivity remains finite.

In Fig. 2a, $\mu_\textrm{2K}$ increases by four orders of magnitude from $n\sim10^{16}$ cm$^{-3}$ to $10^{17}$ cm$^{-3}$, and levels off above $n^*$. Similar trends are observed for RRR and $\rho_\textrm{2K}$ in Fig. 2b and c. Note that $\mu_\textrm{300K}$ and $\rho_\textrm{300K}$ exhibit a slow and smooth evolution with $n$ across $n^*$. The electronic transport is dominated by inelastic electron-phonon scattering near room-$T$ in these cases. In general, this phenomenon was considered to be related to the localization effects in disordered systems, as extensively studied in doped semiconductors~\cite{Spinelli2010}, disordered metals~\cite{Vandendries1981}, and topological materials~\cite{Lu2016}.

Anderson localization describes the localization of a single electron wavefunction in the presence of strong disorder. The envelope of the wavefunction decays exponentially, characterized by a short localization length on the order of the lattice constant\cite{Anderson1958}. While in R-II, we argue that the localization effect is weak and the electrons remain mobile, i.e., the wavefunction is spatially extended. In other words, R-II is seen to be on the metallic side of MIT from the following observations. First, $n$ has a small variation with $T$, indicating that few mobile electrons are frozen at low-$T$. Second, the zero-$T$ conductivity discussed below appears to be finite, which is an alternative definition of metallicity, nevertheless more inclusive than the one mentioned above.

In Fig. 3, we plot the conductivity versus $T$ at low-$T$ for S4, S5, and S6. The data is compared with the formula expressing the combination of the electron-electron interaction and the weak localization effect (WL) in 3D disordered systems\cite{Spinelli2010,Lee1985,Altshuler1980}:

\begin{equation}
\sigma=\sigma_0+a\sqrt{T}+bT^{p/2}
\label{Eq1}
\end{equation}

where $\sigma_0$ is the residual conductivity at zero-$T$, the second term relates to Coulomb interaction effects, and the third term is the 3D WL. The WL effect originates from the constructive interference of a pair of electron wavefunctions, traversing in opposite directions, in self-intersecting scattering paths. $p$ is the power factor in $\tau_\phi^{-1}\propto T^p$. $\tau_\phi$ is the dephasing time of the quantum interference associated with the inelastic scattering of the electrons. $p$ depends on the scattering mechanisms, with values of $\frac{3}{2}$ and two respectively for electron-electron scattering in dirty and clean limits and three for electron-phonon scattering. We observe that: $\sigma_0$ is finite from extrapolation of the conductivity to zero-$T$; both the Coulomb interaction and WL effects play substantial roles (see the figure caption for the details of the fits used); $p$ has values of 3.7, 4.2, and 3 for S4, S5, and S6 respectively. Therefore, our data loosely suggest that electron-phonon scattering is more likely to be the leading mechanism suppressing WL effects than electron-electron scattering.

\begin{figure}
  \includegraphics[width=8cm]{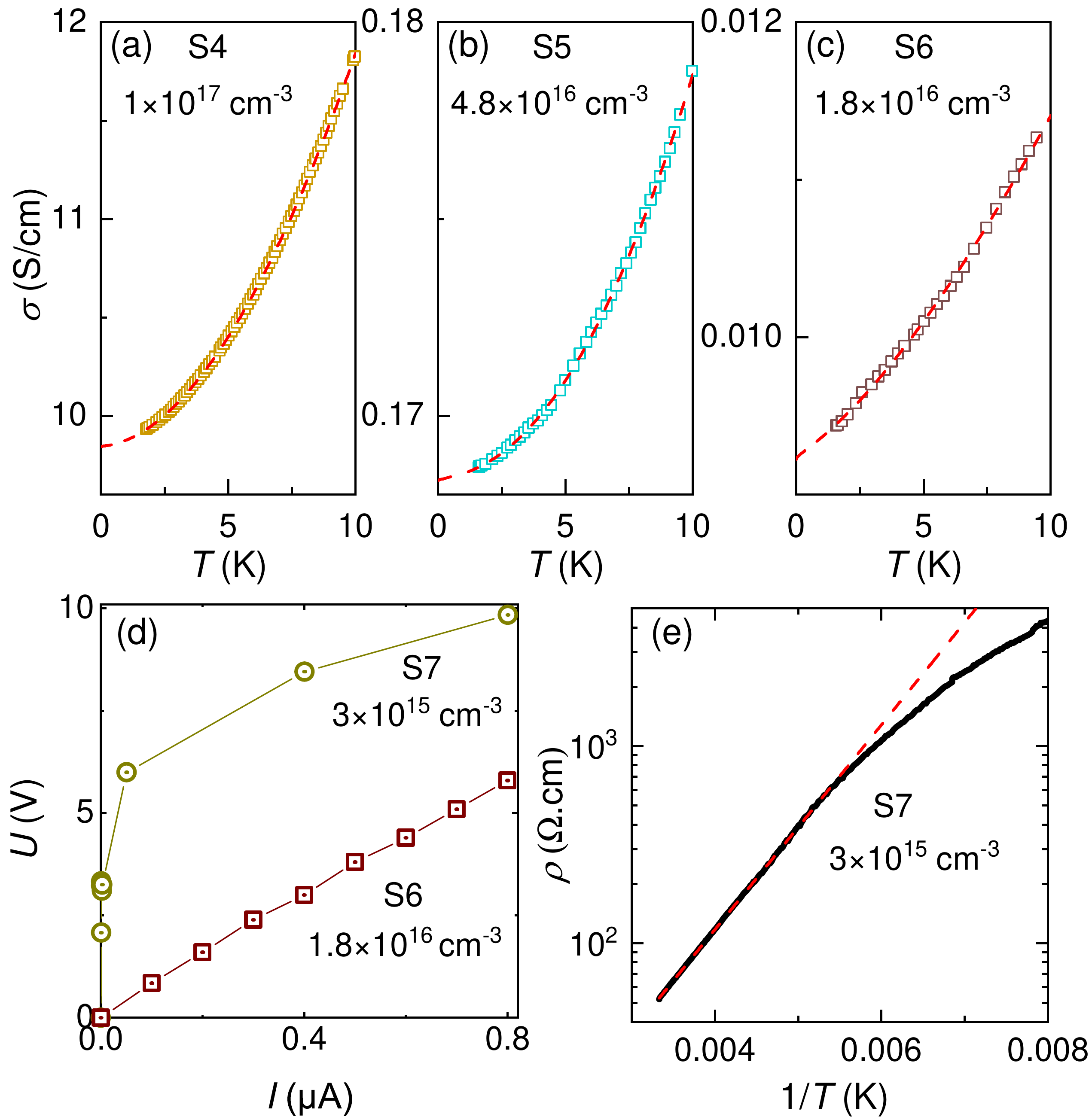}
  \caption{Detailed analysis of transport behaviors. a)-c): Temperature dependent conductivity at low-$T$. The dashed lines are fit to Eq. 1. a)  for S5 with the extracted parameters $\sigma_0=9.9$ S/cm, $a=4.9\times10^{-3}$ S/cm/$\sqrt{\textrm{K}}$, $b=0.027$ S/cm/K$^{p/2}$, and $p=3.7$; b) is for S6 with $\sigma_0=0.168$ S/cm, $a=1.1\times10^{-4}$ S/cm/$\sqrt{\textrm{K}}$, $b=7.6\times10^{-5}$ S/cm/K$^{p/2}$, and $p=4.2$; c) is for S7 with $\sigma_0=9.2\times10^{-3}$ S/cm, $a=7.4\times10^{-5}$ S/cm/$\sqrt{\textrm{K}}$, $b=6.9\times10^{-5}$ S/cm/K$^{p/2}$, and $p=3$. d) $I-V$ curves measured at low-$T$ for S6 and S7. e) $\rho$ versus $1/T$ in a semi-log plot. The dashed line is a fit to $ \rho=\rho_0e^{\frac{\triangle}{k_\textrm{B}T}}$. }
\end{figure}

Let's now turn to R-III (only S7 is included), where MIT occurs. As seen in Fig. 1a, $\rho$ for S7 increases monotonically with the reduction of $T$, and diverges at low temperatures. In Fig. 1b, $n$ declines sharply when decreasing $T$, indicating the immobilization of electrons at low-$T$. The $I-V$ curve, shown in Fig. 3d, is clearly non-ohmic for S7, in contrast with the ohmic behavior shown by S6. All these observations point to an insulating ground state for S7.

Consequently, the transport should be dominated by thermally activated electrons. $\rho$, expressed by $ \rho=\rho_0e^{\frac{\triangle}{k_\textrm{B}T}}$, increases exponentially with decreasing $T$. $\rho_0$ is a constant, $k_\textrm{B}$ is the Boltzmann constant, and $\triangle$ is the activation gap. Correspondingly, a semi-log plot of $\rho$ versus $\frac{1}{T}$ is shown in Fig. 3e for S7. We resolve that $\rho$ follows an activated behavior at high-$T$ with an activation gap of $108$ $\textrm{meV}$, smaller than the indirect band gap of $800$ $\textrm{meV}$. This implies that the shallow impurity levels lie close to, but below, the minimum of the conducting band. This is in contrast to the prediction from DFT calculations~\cite{YanBH2018}.

\begin{figure}
  \includegraphics[width=8.5cm]{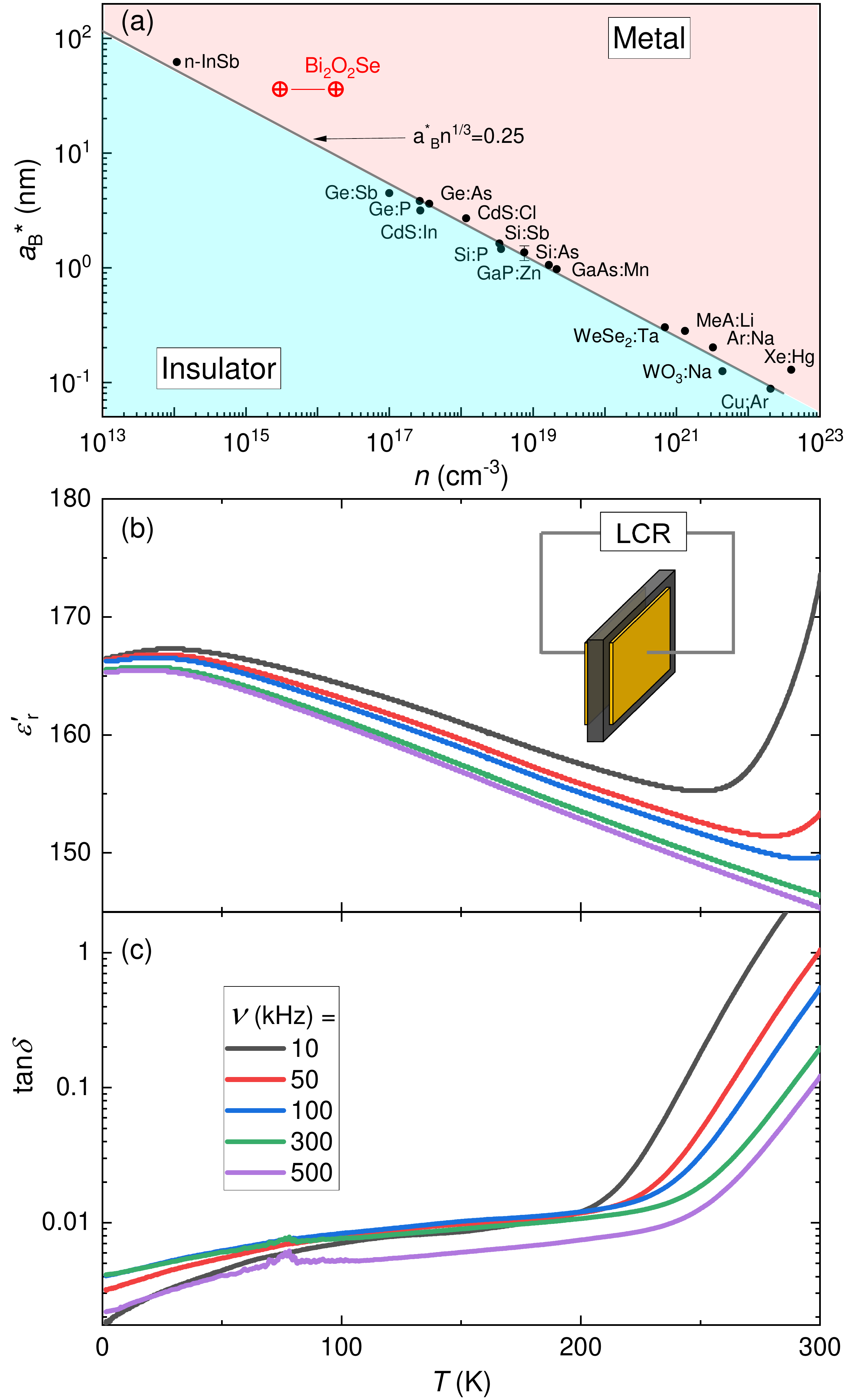}
  \caption{Dielectric measurements of BOS. a): The relation between the critical carrier concentration of MIT ($n_\textrm{c}$) and the effective Bohr radius ($a_\textrm{B}^*$). The diagonal line, from Eq. 2, separates the insulating and metallic phase. The data for doped semiconductors is from ref.~\cite{Edwards1995}. b) and c): Temperature dependence of the permittivity for an insulating sample. b) is the real part of the relative permittivity $\epsilon'_\textrm{r}$. c) is the dissipation factor calculated from $\textrm{tan}\delta=\frac{\epsilon''}{\epsilon'}$, where $\epsilon''$ is the imaginary permittivity.}
\end{figure}

\section{Discussion}\label{sec:4}
We have reached the first main outcome of the paper: MIT occurs at extraordinarily low carrier concentrations, with ($n_\textrm{c}$) between $3\times10^{15}$ cm$^{-3}$ and $1.8\times10^{16}$ cm$^{-3}$, that is, only a single electron is introduced per million unit cells. This is two orders of magnitude lower than the $n_\textrm{c}$ of doped germanium ($n_\textrm{c}\approx 3\times10^{17}$ cm$^{-3}$) and silicon ($n_\textrm{c}\approx 5\times10^{18}$)~\cite{Mott1990}. As a result, negligibly slight nonstoichiometry introduced by crystal defects such as Se or O vacancies, or Se-Bi anti-sites~\cite{YanBH2018,Li2018} can induce MIT. That is why the previously reported samples were metallic. We are still faced with a second issue: why is the critical concentration of MIT in Bi$_2$O$_2$Se so low?

More than half a century ago, Mott proposed a model illustrating the essentiality of electron-electron interactions in the physics underlying MIT in a system~\cite{Mott1956,Mott1961}. It was then summarized as Mott's criterion for MIT~\cite{Mott1990}, expressed by

\begin{equation}
a_\textrm{B}^*n^{1/3}\approx0.25
\label{Eq2}
\end{equation}

where $n$ is the charge carrier density and $a_\textrm{B}^*$ is the effective Bohr radius. Eq. 2 can be achieved by comparing two length scales~\cite{Behnia2015}: $a_\textrm{B}^*$ and the Thomas-Fermi screening length $r_\textrm{TF}$.

$a_\textrm{B}^*$ is the counterpart of the Bohr radius ($a_\textrm{B}$) in a semiconductor. It is defined by $a_\textrm{B}^*=\frac{\epsilon_\textrm{r}}{m^*}a_\textrm{B}$, where $\epsilon_\textrm{r}$ is the relative permittivity and $m^*$ is the effective electron mass in the semiconductor. It quantifies the length scale of the Coulomb field induced by an ionized dopant in a semiconductor.

The second length is the Thomas-Fermi screening length $r_\textrm{TF}$, which characterizes the exponential decay of a screened Coulomb potential in a metal ($V(r)=\frac{e}{4\pi\epsilon r}e^{-r/r_\textrm{TF}}$). The expression for $r_\textrm{TF}$ is as follows~\cite{Ziman1960,Ashcroft1976}:

\begin{equation}
r_\textrm{TF}=\sqrt{\frac{\epsilon}{e^2N(\varepsilon_\textrm{F})}}=\sqrt{\frac{\pi a_\textrm{B}^*}{4k_\textrm{F}}}
\label{Eq3}
\end{equation}

where $N(\varepsilon_\textrm{F})$ is the density of states at the Fermi energy ($\varepsilon_\textrm{F}$) and $k_\textrm{F}$ is the Fermi wave vector.

When $r_\textrm{TF}$ exceeds $a_\textrm{B}^*$, the screening of the Coulomb interaction in the metal extends over the size of the potential well in the insulator (set by $a_\textrm{B}^*$). In this case, the electrons will be trapped in adjacent wells, making the Fermi sea unstable~\cite{Ashcroft1993}. MIT occurs at the critical value:

\begin{equation}
r_\textrm{TF}=a_\textrm{B}^*
\end{equation}

Eq. 4 is an alternative version of Mott's criterion and is equivalent to Eq. 2 by considering an isotropic system, in which $k_\textrm{F}=(3\pi^2n)^{1/3}$.

Note that $a_\textrm{B}^*$ also quantifies the mean radius of the electron orbital around an ionized dopant in a semiconductor. Thus, a qualitative but more straightforward interpretation of Eq. 2 can be that, when the spatial scope of electron motion ($a_\textrm{B}^*$) becomes comparable to their average separation ($n^{-1/3}$), the electron wavefunctions start to overlap. They form a dispersive electronic band from which MIT emerges. This simple scenario has been verified in a variety of insulators, as seen in Fig. 4a~\cite{Mott1990,Edwards1995}. Fig. 4a plots $a_\textrm{B}^*$ versus the critical density $n_\textrm{c}$ for different systems (including doped silicon and germanium). Most systems fall on the diagonal line defined by Eq. 2, which separates the metallic and insulating phases.

To justify its application to BOS, we measured the permittivity in a low-loss insulating sample. Such a measurement was unpractical with previously reported metallic samples. The relative permittivity $\epsilon_\textrm{r}$ along the c-axis, and the dissipation factor ($\textrm{tan}\delta$), as a function of $T$ (measured at different frequencies) are presented in Fig. 4b and c. For low frequencies, $\epsilon_\textrm{r}$ curves upward near room-$T$, accompanied by an enhancement of $\textrm{tan}\delta$. This behavior is akin to that observed in EuTiO$_3$~\cite{Engelmayer2019}, where it was attributed to Schottky-type depletion layers formed at the contact interfaces. The intrinsic $\epsilon_\textrm{r}$, obtained after eliminating the extrinsic contributions at high frequencies, is surprisingly high, amounting to 145 at room-$T$ and increasing slightly to 165 at 2 K. This is one or two orders of magnitude higher than most semiconductors, e.g., SiO$_2$ ($\epsilon_\textrm{r}\approx3.9$), Al$_2$O$_3$ ($\epsilon_\textrm{r}\approx7.8$), and HfO$_2$ ($\epsilon_\textrm{r}\approx25$).

A handful of materials dubbed quantum paraelectrics show huge permittivity due to the softening of optical phonons, e.g., SrTiO$_3$ as seen in Table II. Nevertheless, we argue that BOS is not a quantum paraelectric. First of all, previous reports suggested that there are no soft-phonons in this system~\cite{Wang2020}. Second, it's unlikely that a quantum paraelectric shows such a slight temperature dependence of the permittivity as that in BOS. The origin of the huge permittivity therefore remains an open question. We conjecture that it is linked to particular phonon dispersions in the BOS. DFT calculations predicted several low-lying optical phonon modes in this system~\cite{Wang2020,Wei2019}. For example, the energy of the lowest transverse optical mode ($\omega_{TO}$) at the $\Gamma$ point is merely 8 $\textrm{meV}$.  Note that ultrathin freestanding BOS nanosheets made by a solution-based method were reported to display ferroelectricity at room-$T$, which is a consequence of spontaneous lattice distortion exclusively in freestanding nanosheets~\cite{Ghosh2019}.

Taking the relative permittivity $\epsilon_\textrm{r}\approx155(10)$ and the effective mass $\bar{m}\approx0.23$~\cite{Wang2020}, we derive $a_\textrm{B}^*$ with a value of $36(2)$ $\textrm{nm}$. $\bar{m}$ is taken as the average mass calculated from $\bar{m}=\sqrt{m_\textrm{a}m_\textrm{b}m_\textrm{c}}$, which is based on the anisotropic ellipsoid Fermi pocket of BOS~\cite{Chen2018,Wang2020}. Consequently, we obtain $a_\textrm{B}^*n_\textrm{c}^{1/3}$ values between 0.5 and 1, higher than the critical value of 0.25. Also, as seen in Fig. 4a, BOS is located above the critical line. Thus, we argue that the prematurity the metallic phase in BOS is a direct consequence of the long effective Bohr radius in the framework of Mott's theory of MIT, which is the second main outcome of this paper.

In Table II, we compare the key parameters of BOS with two other doped semiconductors with dilute metallicity. For SrTiO$_3$, a quantum paraelectric, the extremely large permittivity plays a vital role in this context~\cite{Muller1979,Lin2013,Bhattacharya2016}. For InSb, a narrow-gap semiconductor, the low $n_\textrm{c}$ mainly arises from the extraordinarily light effective mass~\cite{Shayegan1988}. Intriguingly, both huge $\epsilon_\textrm{r}$ and light $m^*$ gives equal contributions in BOS.

\begin{table}
\centering
  \caption{Comparison of parameters between Bi$_2$O$_2$Se, SrTiO$_3$ and InSb. $\epsilon_\textrm{r}$ is the relative permittivity, $m^*$ is the effective mass, $a_\textrm{B}^*$ the effective Bohr radius and $n_\textrm{c}$ is the experimental critical concentration of MIT.}

\begin{tabular}{|c|c|c|c|}
  \hline
  & SrTiO$_3$ & InSb~\cite{Shayegan1988} & Bi$_2$O$_2$Se \\

  \hline

  $\epsilon_\textrm{r}$ & 20000~\cite{Muller1979} & 16 & 155(10) \\

  \hline
  $m^*$ ($m_\textrm{e}$) & 1.8~\cite{Lin2013} & 0.014 & 0.23 \\

  \hline
  $a_\textrm{B}^*$ (nm) & 600 & 60 & 36(2) \\

  \hline
  $n_\textrm{c}$ (cm$^{-3}$) & $10^{15}$~\cite{Bhattacharya2016} & $10^{14}$ & $10^{15}-10^{16}$ \\
  \hline

  \end{tabular}
  \label{Tab2}
\end{table}

More stringently, we found that BOS does not exactly fall on the line in Fig. 4a and that $a_\textrm{B}^*n_\textrm{c}^{1/3}$ is at least twice the critical Mott value. There are several possible explanations for the mismatch. First of all, BOS is an anisotropic semiconductor, while Eq. 2 is obtained based on the assumption of an isotropic system. Though we have used the average electron mass ($\bar{m}$) and Fermi wave vector ($k_\textrm{F}$), the permittivity ($\epsilon$) used in the calculation is anisotropic along the c-axis. The mismatch could be reduced if we properly consider the anisotropy of the system. The second explanation is associated with inhomogeneity and is more trivial in nature. In such a dilute limit, the spatial variation of carrier concentrations induced by the random distribution of defects is unavoidable.

\begin{figure}
  \includegraphics[width=8cm]{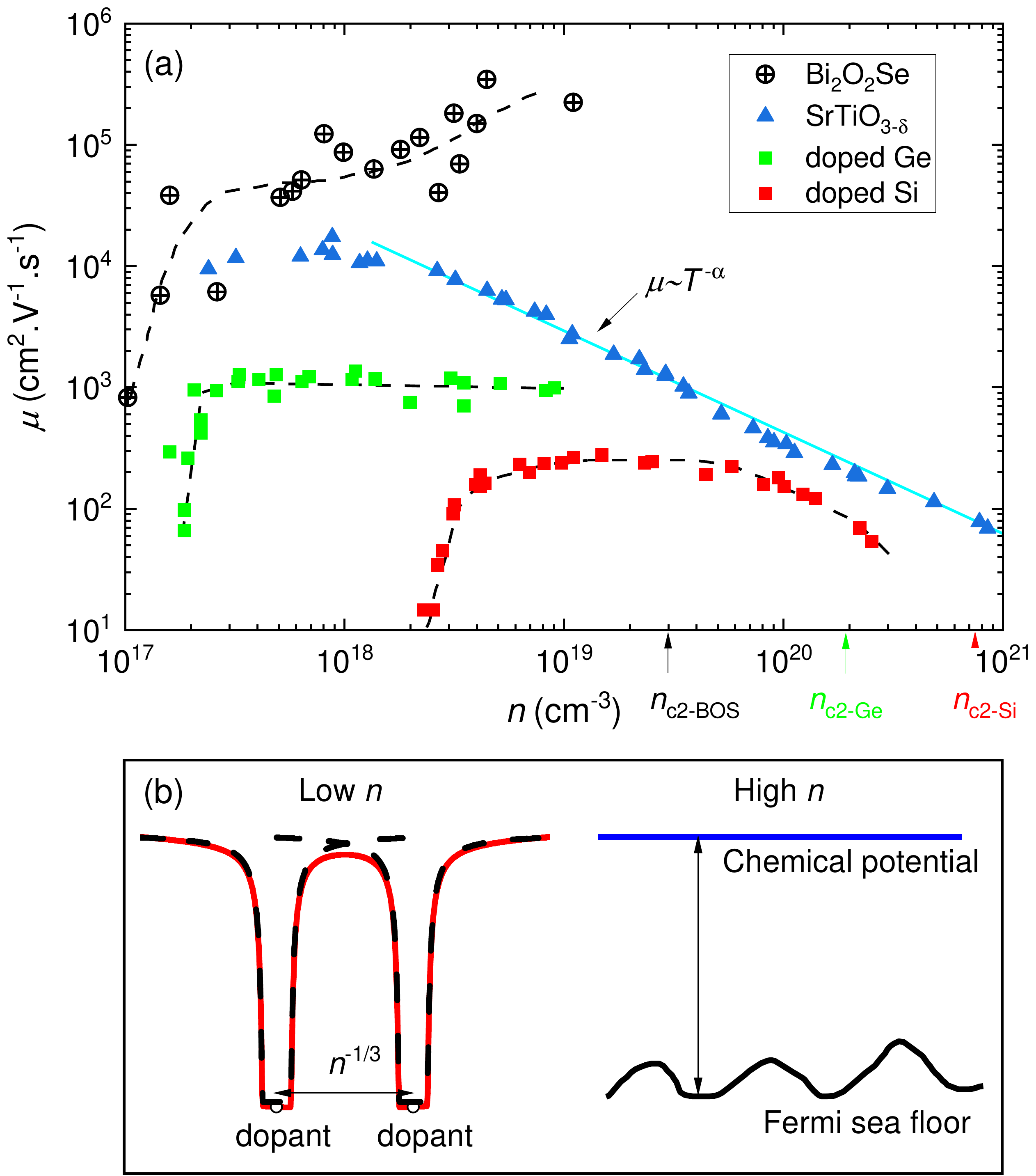}
  \caption{Low-$T$ mobility and potential profiles in real space. a) Carrier dependence of $\mu$ for BOS, compared with SrTiO$_{3-\delta}$~\cite{Wang2019}, doped Si~\cite{Yamanouchi1967} and doped Ge~\cite{Katz1965,Sasaki1975,Watanabe1998}. The dashed lines are guides for the eye. The light blue line marks the power-law behavior. $n_\textrm{c2}$ of BOS, Ge, and Si are marked by arrows. b) Potential profiles for two $n$-regimes. Left: the low-$n$ regime. The dashed curves represent single-dopant potential wells. The red curve summarizes the potential of two wells. The total potential profile is not altered significantly if the two dopants are distant. Right: the high-$n$ regime. The total potential profile forms the Fermi seafloor, whose roughness reflects the randomness in the distribution of dopants. The chemical potential is proportional to the number of carriers. }
\end{figure}

The last remaining question to address is why the low-$T$ electron mobility is quite high. From a simplistic viewpoint, it would be a consequence of the screening of ionized impurity scattering due to the enhanced permittivity. In a number of semiconductors, the mobility dominated by ionized impurity scattering follows a power-law behavior of $n$~\cite{Noguchi1980,Minami1992}: $\mu\sim n^{-\alpha}$, where $\alpha$ is a positive constant. However, in Fig. 2a and Fig. 5a, we see that the $\mu$ of BOS exhibits a slight increase with $n$ above $10^{17} \textrm{cm}^{-3}$, in stark contrast with the power-law behavior of SrTiO$_{3-\delta}$~\cite{Wang2019}.

Behnia~\cite{Behnia2015} argued that the carrier density dependence of $\mu$ is related to the local potential landscape in a shallow Fermi sea. The power-law behavior emerges in regimes where the local potential is composed of the contribution from numerous potential wells, as seen in the right of Fig. 5b. A deviation from the power-law behavior may occur when the potential wells become distant from each other, as seen in the left of Fig. 5b. There is a threshold carrier density $n_\textrm{c2}$ set by~\cite{Behnia2015}:

\begin{equation}
 a_\textrm{B}^*a_0 (n_\textrm{c2})^{2/3}=(\frac{9}{\pi})^{1/3}
 \label{Eq5}
\end{equation}

where $a_0$ is the lattice constant. This simple picture was justified in doped silicon~\cite{Yamanouchi1967} and germanium~\cite{Katz1965,Sasaki1975,Watanabe1998} as seen in Fig. 5a, in which $\mu$ levels off in a range of $n$ below $n_\textrm{c2}$.

Given $a_\textrm{B}^*$ and the in-plane lattice constant ($a_0=3.88$ $\textrm{{\AA}}$) of BOS, we obtain $n_\textrm{c2}$ amounting to $3\times10^{19}$ $\textrm{cm}^{-3}$ beyond our range of study. This may explain the non-power-law behavior. However, the slight increase of $\mu$ with $n$ remains unexplained. We theorize that it may be related to the extent of the homogeneity in the spatial distribution of dopants~\cite{Behnia2015}. In high-$\mu$ BOS samples, the dopants are mainly Se-vacancies ~\cite{YanBH2018}. The higher $\mu$ at higher doping levels may imply that the greater the number of Se-vacancies, the more homogeneous their distribution.

\section{Conclusions}\label{sec:5}

In summary, we observe that MIT emerges at extremely low carrier concentrations in BOS. According to Mott, it is owing to the long effective Bohr radius, stemming from the combination of the surprisingly high $\epsilon_\textrm{r}$ and light $m^*$. The huge $\epsilon_\textrm{r}$ is expected to generate strong Coulomb screening between the electrons and ionized defects/traps. The unexpectedly high $\mu$ at low-$T$ is possibly a consequence of the screening of ionized impurity scattering.

In addition, the results have strong implications for studying the functionality of the semiconductor. For optoelectronics, the screening of trapped centers for photon-excited electrons might be one of the important factors contributing to the record-breaking optoelectronic performance of BOS. These findings may also help us to understand the electron dynamics in the performance of field-effect transistors based on BOS.

\textbf{Acknowledgments}
We acknowledge Kamran Behnia and Wenbin Li for stimulating discussion. This research was supported by the National Natural Science Foundation of China via Project 11904294, Zhejiang Provincial Natural Science Foundation of China under Grant No. LQ19A040005 and the foundation of Westlake Multidisciplinary Research Initiative Center (MRIC)(Grant No. MRIC20200402). We thank the support provided by Dr. Chao Zhang from Instrumentation and Service Center for Physical Sciences at Westlake University.

\textbf{Conflict of Interest}

The authors declare that they have no conflict of interest.








\end{document}